\documentstyle{article}

\catcode`\@=11
\def\relaxnext@{\let\next\relax}
\font\tenmsy=msym10 scaled\magstep1
\font\sevenmsy=msym7 scaled\magstep1
\font\fivemsy=msym5  scaled\magstep1
\newfam\msyfam
\textfont\msyfam=\tenmsy
\scriptfont\msyfam=\sevenmsy
\scriptscriptfont\msyfam=\fivemsy
\font\teneuf=eufm10 scaled\magstep1
\font\seveneuf=eufm7 scaled\magstep1
\font\fiveeuf=eufm5 scaled\magstep1
\newfam\euffam
\textfont\euffam=\teneuf
\scriptfont\euffam=\seveneuf
\scriptscriptfont\euffam=\fiveeuf
\def\frak{\relaxnext@\ifmmode\let\next\frak@\else
 \def\next{\Err@{Use \string\frak\space only in math mode}}\fi\next}
\def\goth{\relaxnext@\ifmmode\let\next\frak@\else
 \def\next{\Err@{Use \string\goth\space only in math mode}}\fi\next}
\def\frak@#1{{\frak@@{#1}}}
\def\frak@@#1{\noaccents@\fam\euffam#1}
\def\Bbb{\relaxnext@\ifmmode\let\next\Bbb@\else
 \def\next{\Err@{Use \string\Bbb\space only in math mode}}\fi\next}
\def\Bbb@#1{{\Bbb@@{#1}}}
\def\Bbb@@#1{\noaccents@\fam\msyfam#1}
\def\accentfam@{7}
\def\noaccents@{\def\accentfam@{0}}
\catcode`\@=\active

\newcommand{\bc}{{\Bbb C}}

\makeatletter
\@addtoreset{equation}{section}
\makeatother

\newtheorem{thm}{Theorem}[section]
\newtheorem{prop}[thm]{Proposition}
\newtheorem{lem}[thm]{Lemma}

\begin{document}
\begin{flushright}
RIMS-1160 \\
hep-th/9708148 \\
Aug. 1997
\end{flushright}

\vspace{36pt}

\begin{center}
\begin{Large}
Form factors of the XXZ model and the affine quantum group symmetry 

\vspace{24pt}
\end{Large}

\begin{large}
Yas-Hiro Quano\raisebox{2mm}{$\star$}

\end{large}
\vspace{24pt}
%\begin{flushleft}
     \it Research Institute for Mathematical Sciences,
          Kyoto University, Kyoto 606-01, Japan  \\
%\end{flushleft}
\vspace{48pt}

\underline{ABSTRACT}
\end{center}

We present new expressions of form factors 
of the XXZ model which satisfy 
Smirnov's three axioms. 
These new form factors are obtained by acting 
the affine quantum group 
$U_q \bigl( \widehat{\frak s \frak l _2}\bigr)$ 
to the known ones obtained in our previous works. 
We also find the relations among all the new and known 
form factors, i.e., all other form factors can be 
expressed as kind of descendents of a special one. 

\vfill
\hrule
 
\vskip 3mm
\begin{small}
 
\noindent\raisebox{2mm}{$\star$} 
Supported by the Grant-in-Aid for Scientific Research
from the Ministry of Education, Science and Culture. 
 
\end{small}
 
\newpage

%\title{Form factors of the XXZ model and 
% the affine quantum group symmetry}

%\author{Yas-Hiro Quano\thanks{
% Supported by the Grant-in-Aid for Scientific Research
% from the Ministry of Education, Science and Culture. }
% }

%\date{\it Research Institute for Mathematical Sciences,
%          Kyoto University, Kyoto 606-01, Japan \hfill}
%\maketitle
%
%\begin{abstract}
%We present new expressions of form factors 
%of the XXZ model which satisfy 
%Smirnov's three axioms. 
%These new form factors are obtained by acting 
%the affine quantum group 
%$U_q \bigl( \widehat{\frak s \frak l _2}\bigr)$ 
%to the known form factors of the model. 
%We also find the relations among all the new and known 
%form factors, i.e., all other form factors are kind of 
%descendent of a specail one. 
%\end{abstract}

\section{Introduction}

In \cite{KMQ} we presented integral formulae 
of solutions to the quantum Knizhnik--Zamolodchikov 
($q$-KZ) equation \cite{FR} of level $0$ 
associated with the vector representation 
of the affine quantum group 
$U_q \bigl( \widehat{\frak s \frak l _2}\bigr)$. 
Those solutions 
satisfy Smirnov's three axioms of form factors \cite{Smbk}. 

Throughout the study of form factors of the sine-Gordon model, 
Smirnov \cite{Smbk} found that 
his three axioms are sufficient conditions of local commutativity of 
local fields of the model. 
Smirnov also constructed the space of local fields 
of the sine-Gordon model \cite{S1}, from the standpoint of 
the form factor bootstrap formalism. 
Smirnov's formulae for form factors of the sine-Gordon 
model are expressed in terms of the deformed Abelian integrals, 
or deformed hyper-elliptic integrals \cite{S2}. 

Babelon, Bernard and Smirnov \cite{BBS1} computed 
form factors of the restricted sine-Gordon model 
at the reflectionless point, by quantizing solitons 
of the model. They also found null vectors of the 
model \cite{BBS2}, which leads to a set of differential 
equations for form factors. 

A form factor is originally defined as 
a matrix element of a local operator. 
In this paper, however, we call any vector valued function 
a `form factor' that satisfies Smirnov's three axioms. 
In this sense, the integral formulae given in \cite{KMQ} 
are form factors of the XXZ model. 
Furthermore, we wish to consider the space of form factors 
of the XXZ model, or solutions of Smirnov's three axioms. 
Our earlier motivation is the question if 
the space of form factors is invariant under the action of 
the affine quantum group 
$U_q \bigl( \widehat{\frak s \frak l _2}\bigr)$, 
which is a symmetry of the XXZ model. 

Let us consider the spin $1/2$ XXZ model with 
the nearest neighbor interaction: 

\begin{equation}
H_{XXZ}=-\frac{1}{2}\sum_{n=-\infty}^{\infty} 
(\sigma_{n+1}^x \sigma_{n}^x + \sigma_{n+1}^y \sigma_{n}^y +
\Delta \sigma_{n+1}^z \sigma_{n}^z), 
\end{equation}
where $\Delta =(q+q^{-1})/2$ and $-1<q<0$. 

Let $V=\Bbb C^2$ be a vector representation of 
the affine quantum group 
$U_q \bigl( \widehat{\frak s \frak l _2}\bigr)$. 
The XXZ Hamiltonian $H_{XXZ}$ formally acts on 
$\cdots \otimes V\otimes V\otimes \cdots$. 
This Hamiltonian commutes with 
$U_q \bigl( \widehat{\frak s \frak l _2}\bigr)$. 
In \cite{XXZ,JM} the space of states $V^{\otimes \infty}$ 
was identified with the tensor product of level $1$ 
highest and level $-1$ lowest representations of 
$U_q \bigl( \widehat{\frak s \frak l _2}\bigr)$. 

Since the XXZ model possesses 
$U_q \bigl( \widehat{\frak s \frak l _2}\bigr)$-symmetry, 
any physical quantities of the model 
are expected to also possess the same symmetry. 
The following question is thus natural: Let 
$G(\zeta_1, \cdots , \zeta_N)$ be a 
$V^{\otimes N}$-valued form factor of the XXZ model. 
Then does $\pi_\zeta (y) G(\zeta )$, where 
$y\in 
U_q \bigl( \widehat{\frak s \frak l _2}\bigr)$, 
again satisfy Smirnov's axioms? 

The answer is as follows: It is not always true that 
$\pi_\zeta (y) G(\zeta )$ solves Smirnov's axioms 
even if $G(\zeta )$ does. However, for a form factor 
that satisfies Smirnov's axioms, there exists 
$y\in U_q \bigl( \widehat{\frak s \frak l _2}\bigr)$ 
such that $\pi_\zeta (y) G(\zeta )$ again satisfies 
the axioms. 

The present paper is organized as follows. 
In section 2 we summarize the results obtained 
in the previous paper \cite{KMQ}. 
In section 3 we present new form factors 
which satisfy Smirnov's three axioms 
by acting the affine quantum group 
$U_q \bigl( \widehat{\frak s \frak l _2}\bigr)$ 
to form factors given in section 2. 
In section 4 we show the relations among 
form factors obtained in sections 2 and 3. 
In section 5 we give some remarks. 

\section{Integral formula of form factors of the XXZ model}

In this section we review Smirnov's three axioms of 
form factors \cite{Smbk} and the integral formula of form factors 
of the XXZ model given in \cite{KMQ}. See 
\cite{JKMQ,KMQ} as for explicit expressions of 
some scalar functions and homogeneous functions below. 

For a fixed complex parameter $q$ such that $0<x=-q<1$, 
let $U$ be the affine quantum group 
$U'_q \bigl(\widehat{\frak s \frak l _2}\bigr)$ 
generated by $e_i, f_i, t_i(i=0,1)$ \cite{J}. 
Set $V=\bc v_+ \oplus \bc v_- $ and 
let 
$(\pi_\zeta, V)$, where $\zeta\in \Bbb C \backslash \{0\}$, 
denote the vector representation of $U$ defined by
\begin{equation}
\begin{array}{l}
\pi_{\zeta} (e_1)(v_+ , v_-)=\zeta(0,v_+), \quad
\pi_{\zeta} (f_1) (v_+ , v_-)=\zeta^{-1}(v_-,0), \quad
\pi_{\zeta} (t_1) (v_+ , v_-)=(q v_+ , q^{-1} v_-), \\
\pi_{\zeta} (e_0)(v_+ , v_-)=\zeta(v_-,0), \quad
\pi_{\zeta} (f_0) (v_+ , v_-)=\zeta^{-1}(0,v_+), \quad
\pi_{\zeta} (t_0) (v_+ , v_-)=(q^{-1} v_+ , q v_-).
\end{array}
\label{eqn:pv}
\end{equation}

Let $R(\zeta), S(\zeta) =S_0(\zeta)R(\zeta) \in 
\mbox{End}(V \otimes V)$ 
be the $R$ and $S$ matrix of the XXZ model, 
where the ratio $S_0(\zeta)$ is a scalar function, 
which satisfy the intertwining property \cite{J}: 

\begin{equation}
X(\zeta_1 /\zeta_2)
(\pi_{\zeta_1} \otimes\pi_{\zeta_2})\circ\Delta(y)=
(\pi_{\zeta_1} \otimes\pi_{\zeta_2})\circ\Delta'(y)
X(\zeta_1 /\zeta_2), 
\label{eqn:intertwine}
\end{equation}

for $X=R$ or $S$; 
$\Delta$ and $\Delta'=\sigma \circ \Delta$ 
are coproducts of $U$. 

For $n\geq 0, l\geq 0, n+l=N$, 
let $V^{(n l)}$ be a subspace of $V^{\otimes N}$ 
such that 
$$
V^{(n l)}
=\bigoplus_{\sum \varepsilon_i =l-n} \Bbb C v_{\varepsilon_1} \otimes
\cdots \otimes v_{\varepsilon_N}. 
$$
Here we set $N$ even such that $n\equiv l$ mod $2$, 
for the simplicity. The odd $N$ cases can be treated 
similarly. 

Let $G^{(n l)}_{\varepsilon}(\zeta_1,\cdots,\zeta_N)
\in V^{(n l)}$ with $\varepsilon =\pm$ 
be a form factor that satisfies the 
following three axioms: 

\noindent{\bf 1. $S$-matrix symmetry}
\begin{equation}
P_{j\,j+1} G_{\varepsilon}^{(n l)} 
(\cdots,\zeta_{j+1},\zeta_j,\cdots) 
\quad =
S_{j\,j+1}(\zeta_j/\zeta_{j+1})G_{\varepsilon}^{(n l)} 
(\cdots,\zeta_j,\zeta_{j+1},\cdots)
\qquad (1\leq j\leq N-1), 
\label{eqn:S-symm} \\
\end{equation}
where 
$P(x\otimes y)=y\otimes x$ for $x,y\in V$. 

\noindent{\bf 2. Deformed cyclicity}
\begin{equation}
P_{12}\cdots P_{N-1 N} G_{\varepsilon}^{(n l)} 
(\zeta_2,\cdots,\zeta_N, \zeta_1 q^{-2})
=
\varepsilon r(\zeta_1) D_1 G_{\varepsilon}^{(n l)} 
(\zeta_1,\cdots,\zeta_N), 
\label{eqn:cyc}\\
\end{equation}
where $r(\zeta)$ is an appropriate scalar function and 
$D_1$ is a diagonal operator of the form 
$D_1=D\otimes 1\otimes \cdots \otimes 1$. 
The LHS of (\ref{eqn:cyc}) is the analytic continuation 
of $P_{12}\cdots P_{N-1 N} G_{\varepsilon}^{(n l)} 
(\zeta_2,\cdots,\zeta_N, \zeta_1)$ in the variable $\zeta_1$. 

\noindent{\bf 3. Annihilation pole condition} 
The $G_{\varepsilon}^{(n l)} (\zeta)$ has simple poles at 
$\zeta_N =\sigma\zeta_{N-1}x^{-1}$ 
with $\sigma=\pm$, and the residue is given by 
\begin{eqnarray}
&& \displaystyle  
{\rm Res}_{\zeta_N =\sigma\zeta_{N-1}x^{-1}} 
G_{\varepsilon}^{(n l)} (\zeta)
\label{eqn:Res} \\
&=& \frac{1}{2}
\displaystyle\left( I
-\varepsilon \sigma^{N+1} 
r(\sigma\zeta_{N-1}x) 
D_N S_{N-1,N-2}(\zeta_{N-1}/\zeta_{N-2}) 
\cdots S_{N-1,1}(\zeta_{N-1}/\zeta_{1}) 
\right)
G_{\varepsilon}^{(n-1 l-1)}(\zeta') \otimes u_{\sigma} , 
\nonumber
\end{eqnarray}
where $(\zeta)=(\zeta_1, \cdots , \zeta_N), 
(\zeta')=(\zeta_1, \cdots , \zeta_{N-2})$, and  
$u_{\sigma} =v_+ \otimes v_- 
           +\sigma v_- \otimes v_+ $. 
Here $D_N$ is of course a diagonal operator 
of the form $D_N =1\otimes \cdots \otimes 1\otimes D$. 

~

{\it Remark 1.} Note that 
the consistency of these three axioms 
implies the relation 
$r(\zeta)r(\sigma\zeta x)=\sigma^N$. 

~

{\it Remark 2.} 
Let $|\mbox{vac}\rangle _i$ ($i=0, 1$) 
be the ground states of the XXZ model, 
where the subscript $i$ signifies the 
boundary condition of the ground state. 
{}From the standpoint of the 
vertex operator formalism, 
$|\mbox{vac}\rangle _i$ 
is the canonical element 
of $V(\Lambda_i)\otimes V(\Lambda_i)^*$, 
where $V(\Lambda_i)$ is the level $1$ 
highest weight module of the affine 
quantum group $U$ \cite{XXZ}. 
Let $\varphi^*(\zeta)$ denote the creation 
operator. 
Then the form factor of the local operator 
${\cal O}$ is given as follows: 
$$
G^{(N)}_i (\zeta_1,\cdots,\zeta_N)
={}_{i}\langle{\rm vac}|
{\cal O}\varphi^*(\zeta_N)\cdots \varphi^*(\zeta_1)
|{\rm vac}\rangle_{i}. 
$$
We set $G^{(N)}_\varepsilon(\zeta)
=G^{(N)}_0 (\zeta)+\varepsilon G^{(N)}_1 (\zeta)$ 
such that the second and third axioms 
(\ref{eqn:cyc}--\ref{eqn:Res}) reduce the closed form 
in terms of 
$G^{(N)}_\varepsilon (\zeta)$. 

~

In \cite{JKMQ,KMQ} we constructed a solution of 
(\ref{eqn:S-symm}--\ref{eqn:Res}) as follows. 
Set $m=n-1$ for $n=l$ and $m=\mbox{min}(n,l)$ 
for $n\neq l$, and set 
$D=D^{(nl)}=q^{-N/2}\mbox{diag}(q^n, q^l)$. 
Let 
$\Delta^{(nl)}(x_1, \cdots , x_m| z_1,\cdots,z_n |
z_{n+1}, \cdots , z_N )$ be 
a homogeneous polynomial of $x$'s and $z$'s, 
antisymmetric with respect to $x$'s and 
symmetric with $z_j$'s($1\leq j \leq n$) and 
$z_i$'s($n+1\leq i \leq N$), respectively. 
For such a polynomial, let us define 
$\langle \Delta^{(nl)} \rangle 
(x_1, \cdots , x_m| \zeta_1,\cdots,\zeta_N ) 
\in V^{\otimes N}$ by 
\begin{eqnarray}
&&\langle \Delta^{(nl)} \rangle 
(x_1, \cdots , x_m| \zeta_1,\cdots,\zeta_N ) ^{-\cdots -+\cdots +} 
=
\displaystyle\Delta^{(nl)} \bigl(
x_1, \cdots , x_m | 
z_1, \cdots , z_n |z_{n+1}, \cdots , z_N \bigr)
\prod_{j=1}^{n} \zeta_j \left( 
\prod_{i=n+1}^{N} \frac{1}{z_i -z_j \tau^2} \right), 
\nonumber \\
&&P_{j\,j+1} \langle \Delta^{(nl)} \rangle 
(x_1, \cdots , x_m | \cdots,\zeta_{j+1},\zeta_j,\cdots) 
=
R_{j\,j+1}(\zeta_j/\zeta_{j+1})
\langle \Delta^{(nl)} \rangle 
(x_1, \cdots , x_m | \cdots,\zeta_j,\zeta_{j+1},\cdots), 
\label{eqn:Rsym}
\end{eqnarray}
where $z_j=\zeta_j^2 (1\leq j\leq N)$. 

Assume that $n\leq l$ for a while. 
Then an integral formula 
that solves all the three axioms 
(\ref{eqn:S-symm}--\ref{eqn:Res}) is given as follows: 

\begin{equation}
G_{\varepsilon}^{(nl)}(\zeta) 
=
\displaystyle 
\frac{G_0^{(N)}(\zeta)}{m!}
\prod_{\mu =1}^{m} \oint_{C} \frac{dx_{\mu }}{2\pi i} 
\Psi_{\varepsilon}^{(m N)}
(x_1 , \cdots , x_m | \zeta_1 , \cdots , \zeta_N ) 
\langle \Delta^{(nl)}  \rangle 
(\zeta_1,\cdots,\zeta_N ), 
\label{eqn:ref_G}
\end{equation}
where $G_0(\zeta)$ is an appropriate scalar function. 

The path of the integral $C=C(z_1, \cdots , z_N)$ 
and the explicit expression 
of the integral kernel 
$\Psi_{\varepsilon}^{(m N)}$ are not important 
in this paper. See \cite{JKMQ,KMQ} as for details. 
Note that (\ref{eqn:Rsym}) ensures the $S$-matrix symmetry 
(\ref{eqn:S-symm}). The second axiom (\ref{eqn:cyc}) 
and the third one (\ref{eqn:Res}) 
imply the transformation properties \cite{JKMQ} 
and the recursion relation \cite{KMQ} of the kernel 
$\Psi_\varepsilon ^{(m N)}(x|\zeta)$, respectively. 

The explicit expression of $\Delta^{(nl)}$ 
is also unimportant in this paper. 
The essential point concerning $\Delta^{(nl)}$ 
is the following recursion relations 
\begin{eqnarray}
&& \Delta^{(nl)}(x_1,\cdots,x_m|z_1, \cdots , z_n |
z_{n+1}, \cdots , z_N)|_{z_N=z_n q^{-2}} 
\label{eqn:Drec} \\
&=& \displaystyle\prod_{\mu=1}^m (x_\mu-z_n q^{-1})
\sum_{\nu=1}^m (-1)^{m+\nu} 
h^{(N-2)}(x_\nu|z_1, \mathop{\hat{\cdots}}^n , z_{N-1})
\Delta^{(n-1 l-1)}(x_1,\mathop{\hat{\cdots}}^\nu,x_m|
z_1, \cdots, z_{n-1}|z_{n+1}, \cdots, z_{N-1}), 
\nonumber
\end{eqnarray}
where $h^{(N)}(x|z_1, \cdots, z_N)$ is a homogeneous 
function of degree $N-1$ \cite{JKMQ}, 
and the degree condition 
\begin{equation}
\mbox{deg} \Delta^{(nl)}=\left( 
\begin{array}{c} m \\ 2 \end{array} \right) +nl-n. 
\label{eqn:Ddeg}
\end{equation}
Note that one can determine $\Delta^{(nl)}$ 
recursively by using (\ref{eqn:Drec}). 
From the antisymmetry with respect to $x$'s, 
$\Delta^{(nl)}$ has the factor 
$\displaystyle \prod_{\mu <\nu} (x_\mu -x_\nu)$. 
Hence $\Delta^{(nl)}
/\displaystyle \prod_{\mu <\nu} (x_\mu -x_\nu)$ is 
a polynomial of degree $nl-n$. 
{}From the symmetry property with 
respect to $z$'s, the recursion relation (\ref{eqn:Drec}) 
gives values of $\Delta^{(nl)}$ at $nl$ points. Thus 
the polynomial $\Delta^{(nl)}$ can be determined 
from the initial conditions 
\begin{equation}
\Delta^{(0l)}=\Delta^{(11)}=1, ~~l>0. 
\label{eqn:Dini}
\end{equation}

\section{Form factors and the action of 
the affine quantum group}

In this section we discuss the transformation 
properties of the form factors given in the last section 
under the action of the affine quantum group 
$U=U'_q \bigl(\widehat{\frak s \frak l _2}\bigr)$. 

For any $y\in U$, the tensor representation 
$(\pi_{(\zeta_1,\cdots,\zeta_N)}(y), V^{\otimes N})$ 
is defined as follows: 
\begin{equation}
 \pi_{(\zeta_1,\cdots,\zeta_N)}(y)
=(\pi_{\zeta_1}\otimes\cdots\otimes\pi_{\zeta_N})\circ
\Delta^{(N-1)}(y). 
\end{equation}
Let us act $\pi_{(\zeta_1,\cdots,\zeta_N)}(y)$ 
to $G_\varepsilon^{(nl)}(\zeta_1, \cdots, \zeta_N)$. 
The action of $t_i$'s are trivial: 
$$
\pi_{\zeta}(t_0)G_\varepsilon^{(nl)}(\zeta)
=q^{n-l}G_\varepsilon^{(nl)}(\zeta), ~~
\pi_{\zeta}(t_1)G_\varepsilon^{(nl)}(\zeta)
=q^{l-n}G_\varepsilon^{(nl)}(\zeta). 
$$

The action of $f_0$ is non-trivial but the result is 
very simple: 
\begin{lem}
\begin{equation}
\pi_{\zeta}(f_0)G_\varepsilon^{(nl)}(\zeta)=0, 
\label{eqn:f0G}
\end{equation}
\end{lem}

{\sl Proof. }~
In order to prove (\ref{eqn:f0G}) 
it is enough to show 
\begin{equation}
\pi_{\zeta}(f_0)\langle \Delta^{(nl)} \rangle(x|\zeta) =0.
\label{eqn:hw}
\end{equation}
The $R$-matrix symmetry (\ref{eqn:Rsym}) of 
$\langle \Delta^{(nl)} \rangle(x|\zeta)$ implies that 
of $\pi_{\zeta}(f_0)\langle \Delta^{(nl)} \rangle(x|\zeta)$ 
from the intertwining property (\ref{eqn:intertwine}). 
The arbitrary component of 
$\pi_{\zeta}(f_0)\langle \Delta^{(nl)} \rangle(x|\zeta)$ 
can be expressed in terms of linear combination of 
the extreme component 
$(\pi_{\zeta}(f_0)\langle \Delta^{(nl)} \rangle
(x|\zeta_{s(1)}, \cdots, \zeta_{s(N)}))^{
-\cdots -+\cdots +}$'s, where 
$s \in \frak S_N$. 
The claim (\ref{eqn:hw}) thus follows 
from that $(\pi_{\zeta}(f_0)\langle \Delta^{(nl)} \rangle(x|\zeta))^{
-\cdots -+\cdots +}$ vanishes. 

Set $\langle \Delta_{(n-1 l+1)}^{(0)} \rangle(x|\zeta)
=\pi_{\zeta}(f_0)\langle \Delta^{(nl)} \rangle(x|\zeta)$.
Then $\Delta_{(n-1 l+1)}^{(0)}(x_1, \cdots, x_m| 
z_1, \cdots, z_{n-1}|z_n, \cdots, z_N)$ is proportional to 
$(\pi_{\zeta}(f_0)\langle \Delta^{(nl)} \rangle(x|\zeta))^{
-\cdots -+\cdots +}$. Thanks to the $R$-matrix symmetry 
we obtain
\begin{eqnarray}
\begin{array}{cl}
&\Delta_{(n-1 l+1)}^{(0)}
(x_1,\cdots,x_m|z_1,\cdots,z_{n-1}|z_n,\cdots,z_N) \\
=&\displaystyle \sum_{k=n}^{N} 
\frac{\prod_{j=1}^{n-1}(z_k-z_j q^{-2})}
{\prod_{\scriptstyle{i=n}\atop \scriptstyle{i\ne k}}^{N} 
(z_i-z_k)q^{-1}}
\Delta^{(n l)}(x_1,\cdots,x_m|z_1,\cdots, z_{n-1},z_k|
z_{n},\mathop{\hat{\cdots}}^{k},z_N), 
\end{array}
\label{eqn:f0}
\end{eqnarray}

Note that the singularity at $z_k=z_i$ in 
the RHS of (\ref{eqn:f0}) is spurious, and 
hence that $\Delta_{(n-1 l+1)}^{(0)}$ 
is a homogeneous polynomial of degree 
$\displaystyle{{m \choose 2}}+(n-1)(l+1)-n$, 
antisymmetric with respect to $x_\mu$'s and symmetric
with respect to
$\{z_1,\cdots,z_{n-1}\}$ and $\{z_{n},\cdots,z_N\}$, respectively.
The recursion relation 
$$
\begin{array}{cl}
&\Delta_{(n-1\,l+1)}^{(0)}
(x_1,\cdots,x_m|z_1, \cdots, z_{n-1}|z_n, \cdots, z_N)|_{
z_N=z_{n-1}q^{-2}} \\
=& \displaystyle\prod_{\mu=1}^m (x_\mu-z_n q^{-1})
\sum_{\nu=1}^m (-1)^{m+\nu} 
h^{(N-2)}(x_\nu|z_1, \mathop{\hat{\cdots}}^n , z_{N-1})
\Delta_{(n-2\, l)}(x_1,\mathop{\hat{\cdots}}^\nu,x_m|
z_1, \cdots, z_{n-2}|z_{n}, \cdots, z_{N-1}), 
\end{array}
$$
is enough to determine $\Delta_{(n-1 l+1)}^{(0)}$ 
recursively. 
{}From the power counting $\Delta_{(0 l+1)}^{(0)}=0$. 
Thus $\Delta_{(n-1 l+1)}^{(0)}=0$, which implies (\ref{eqn:hw}). 
~~~~$\Box$

~

From now on, we wish to consider 
$\pi_\zeta(y)G^{(nl)}(\zeta)$ for $y\in U$. 
For that purpose, let us list the following formulae 
for 
$\langle \Delta_{(n+1 l-1)}^{(0)} \rangle(x|\zeta)
=\pi_{\zeta}(e_0)\langle \Delta^{(nl)} \rangle(x|\zeta)$, 
$\langle \Delta_{(n-1 l+1)}^{(1)} \rangle(x|\zeta)
=\pi_{\zeta}(e_1)\langle \Delta^{(nl)} \rangle(x|\zeta)$, 
and 
$\langle \Delta_{(n+1 l-1)}^{(1)} \rangle(x|\zeta)
=\pi_{\zeta}(f_1)\langle \Delta^{(nl)} \rangle(x|\zeta)$: 

\begin{equation}
\begin{array}{cl}
&\Delta_{(n+1 l-1)}^{(0)}
(x_1,\cdots,x_m|z_1,\cdots,z_{n+1}|z_{n+2},\cdots,z_N) \\
=&\displaystyle\sum_{k=1}^{n+1}
\frac{\prod_{i=n+2}^{N}(z_i-z_k q^{-2})}
{\prod_{\scriptstyle{j=1}\atop \scriptstyle{j\ne k}}^{n+1} 
(z_k-z_j)q^{-1}}
\Delta^{(n l)}(x_1,\cdots,x_m|z_1,\mathop{\hat{\cdots}}^{k},z_{n+1}|
z_k, z_{n+2},\cdots,z_N), 
\end{array}
\label{eqn:e0}
\end{equation}
\begin{equation}
\begin{array}{cl}
&\Delta_{(n-1 l+1)}^{(1)}
(x_1,\cdots,x_m|z_1,\cdots,z_{n-1}|z_n,\cdots,z_N) \\
=&\displaystyle q^{l-n+1}\sum_{k=n}^{N} z_k
\frac{\prod_{j=1}^{n-1}(z_k-z_j q^{-2})}
{\prod_{\scriptstyle{i=n}\atop \scriptstyle{i\ne k}}^{N} 
(z_i-z_k)q^{-1}}
\Delta^{(n l)}(x_1,\cdots,x_m|z_1,\cdots, z_{n-1},z_k|
z_{n},\mathop{\hat{\cdots}}^{k},z_N), 
\end{array}
\label{eqn:e1}
\end{equation}
\begin{equation}
\begin{array}{cl}
&\Delta_{(n+1 l-1)}^{(1)}
(x_1,\cdots,x_m|z_1,\cdots,z_{n+1}|z_n,\cdots,z_N) \\
=&\displaystyle q^{n-l+1} \sum_{k=1}^{n+1} z_k^{-1}
\frac{\prod_{i=n+2}^{N}(z_i-z_k q^{-2})}
{\prod_{\scriptstyle{j=1}\atop \scriptstyle{j\ne k}}^{n+1} 
(z_k-z_j)q^{-1}}
\Delta^{(n l)}(x_1,\cdots,x_m|z_1,\mathop{\hat{\cdots}}^{k}
,z_{n+1}|z_k, z_{n+2},\cdots,z_N). 
\end{array}
\label{eqn:f1}
\end{equation}
The expressions (\ref{eqn:e0}--\ref{eqn:f1}) can be 
proved in a similar manner as (\ref{eqn:f0}) is shown. 

Let $A_0(\zeta)=G_\varepsilon^{(nl)}(\zeta)$, and 
$A_j(\zeta)=\pi_\zeta(e_0)A_{j-1}(\zeta)$, where 
$1\leq j$. Then $A_j(\zeta)=0$ for $j>l-n$, and 
$A_{j-1}(\zeta)=\mbox{const.}
\pi_\zeta(e_0)A_{j}(\zeta)$ for 
$1\leq j \leq l-n$. These $(l-n+1)$ 
$\left\{A_j(\zeta)\right\}_{0\leq j\leq l-n}$ 
form a multiplet. Now the following 
natural question arises: Do all $A_j(\zeta)$'s satisfy 
the three axioms (\ref{eqn:S-symm}--\ref{eqn:Res}) 
for a suitable choice of the diagonal operator $D$? 

The $S$-matrix symmetry is apparently satisfied 
by any $A_j(\zeta)$. The second and third axioms 
are unfortunately invalid unless $n=l$. (Since 
$\pi_\zeta(f_0)G_\varepsilon^{(nn)}(\zeta)=0=
\pi_\zeta(e_0)G_\varepsilon^{(nn)}(\zeta)$, the case $n=l$ is 
trivial.) 

However, we can consider if 
$\pi_\zeta(y)G_\varepsilon^{(nl)}(\zeta)$ 
do satisfy the three axioms, where $y=e_1$ or $f_1$, 
because $\pi_\zeta(y)G_\varepsilon^{(nl)}(\zeta)$ 
for any $y\in U$ 
always satisfy the first axiom (\ref{eqn:S-symm}). 
We do not have to restrict ourselves to the case 
$y=e_0$. Actually, 
$\tilde{G}_\varepsilon^{(n+1 l-1)}:
=\pi_\zeta(f_1)G_\varepsilon^{(nl)}(\zeta)$ 
solves all the three axioms for 
$D=q^{-N/2}\mbox{diag}(q^{n-1}, q^{l+1})$: 

\begin{thm}
The vector 
$\tilde{G}_\varepsilon^{(n+1 l-1)}
:=\pi_\zeta(f_1)G_\varepsilon^{(nl)}(\zeta)$ 
satisfies (\ref{eqn:S-symm}--\ref{eqn:Res}) 
when we set the diagonal operator $D=D^{(n-1 l+1)}$. 
\end{thm}

{\sl Proof. }~ 
The proof is straightforward. 
By noticing that  
the LHS of (\ref{eqn:cyc}) should be interpreted 
as the analytic continuation in the variable $\zeta_1$, 
we have 
\begin{equation}
\begin{array}{cl}
&P_{12}\cdots P_{N-1 N}
\pi_{(\zeta_2, \cdots, \zeta_N, \zeta_1 q^{-2})}(f_1) 
G_\varepsilon^{(nl)}(\zeta_2, \cdots, \zeta_N, \zeta_1 q^{-2}) \\
=&(\pi_{(\zeta_1 q^{-2}, \zeta_2, \cdots, \zeta_N)}
\circ \Delta' (f_1)) P_{12}\cdots P_{N-1 N}
G_\varepsilon^{(nl)}(\zeta_2, \cdots, \zeta_N, \zeta_1 q^{-2}) \\
=&\varepsilon r(\zeta_1)
(\pi_{\zeta_1 q^{-2}}(f_1) \otimes 
\pi_{(\zeta_2, \cdots, \zeta_N)}(1)
+\pi_{\zeta_1 q^{-2}}(t_1^{-1}) \otimes 
\pi_{(\zeta_2, \cdots, \zeta_N)}(f_1))
D^{(nl)}_1 G_\varepsilon^{(nl)}(\zeta_1, \cdots, \zeta_N) \\
=&\varepsilon r(\zeta_1)
D^{(nl)}_1 (q^{n-l} \pi_{\zeta_1}(t_1^{-1}f_1 t_1) \otimes  
\pi_{(\zeta_2, \cdots, \zeta_N)}(t_1^{-1}t_1)
+\pi_{\zeta_1}(t_1^{-1}) \otimes 
\pi_{(\zeta_2, \cdots, \zeta_N)}(f_1))
G_\varepsilon^{(nl)}(\zeta_1, \cdots, \zeta_N) \\
=&\varepsilon r(\zeta_1)(D^{(nl)}t_1^{-1})_1 
(\pi_{\zeta_1}(f_1) \otimes  
\pi_{(\zeta_2, \cdots, \zeta_N)}(t_1^{-1})
+\pi_{\zeta_1}(1) \otimes 
\pi_{(\zeta_2, \cdots, \zeta_N)}(f_1))
G_\varepsilon^{(nl)}(\zeta_1, \cdots, \zeta_N) \\
=&\varepsilon r(\zeta_1)(D^{(nl)}t_1^{-1})_1 
\tilde{G}_\varepsilon^{(n+1 l-1)}(\zeta). 
\end{array}
\label{eqn:actf1}
\end{equation}
Thus the second axiom is proved. 

The third axiom 
for $\tilde{G}_\varepsilon^{(n+1 l-1)}(\zeta)$ can be proved 
as follows. Since the action of 
$\pi_{\zeta}(f_1)$ produces no further singularity 
of form factors, we have 
$$
{\rm Res}_{\zeta_N =\sigma \zeta_{N-1}x^{-1}}
\tilde{G}_\varepsilon^{(n+1 l-1)}(\zeta)
=\pi_{(\zeta',\zeta_{N-1},\sigma \zeta_{N-1}x^{-1})}(f_1)
{\rm Res}_{\zeta_N =\sigma \zeta_{N-1}x^{-1}}
G_\varepsilon^{(n+1 l-1)}(\zeta). 
$$
Note that 
$\pi_{(\zeta_{N-1},\sigma \zeta_{N-1}x^{-1})}(f_1)
u_\sigma =0$. 
We also notice that 
\begin{equation}
\begin{array}{cl}
&\pi_{(\zeta',\zeta_{N-1},\sigma \zeta_{N-1}x^{-1})}(f_1)
S_{N-1,N-2}(\zeta_{N-1}/\zeta_{N-2})\cdots 
S_{N-1,1}(\zeta_{N-1}/\zeta_{1}) \\
=&
S_{N-1,N-2}(\zeta_{N-1}/\zeta_{N-2})\cdots 
S_{N-1,1}(\zeta_{N-1}/\zeta_{1}) 
(\pi_\zeta'(t_1^{-1})\otimes \pi_{\zeta_{N-1}}(f_1)\otimes 
\pi_{\zeta_N}(t^{-1}_1) \\
&+
\pi_\zeta'(f_1)\otimes \pi_{\zeta_{N-1}}(1)\otimes 
\pi_{\zeta_N}(t^{-1}_1)+
\pi_\zeta'(1)\otimes \pi_{\zeta_{N-1}}(1)\otimes 
\pi_{\zeta_N}(f_1)), \label{eqn:f1Res}
\end{array}
\end{equation}
and that the first and the third term of the RHS of 
(\ref{eqn:f1Res}) cancel when they act on 
$G_\varepsilon^{(n-1 l-1)}(\zeta')\otimes u_\sigma$. 
Thus we obtain 
$$
\begin{array}{cl}
&{\rm Res}_{\zeta_N =\sigma \zeta_{N-1}x^{-1}}
\tilde{G}_\varepsilon^{(n+1 l-1)}(\zeta) \\
=&
\frac{1}{2}
\displaystyle\left( I
-\varepsilon \sigma^{N+1} 
r(\sigma\zeta_{N-1}x) 
(t_1^{-1}D^{(nl)})_N S_{N-1,N-2}(\zeta_{N-1}/\zeta_{N-2}) 
\cdots S_{N-1,1}(\zeta_{N-1}/\zeta_{1}) 
\right)
\tilde{G}_{\varepsilon}^{(n-1 l-1)}(\zeta') \otimes u_{\sigma}, 
\end{array}
$$
that implies the third axiom with 
$D=D^{(n-1 l+1)}$. ~~~~$\Box$

~

Note that the diagonal operator for 
$\tilde{G}_\varepsilon^{(n+1 l-1)}(\zeta)$ 
is $D^{(n-1 l+1)}$ but not 
$D^{(n+1 l-1)}$, so that 
$\pi_\zeta(f_1)\tilde{G}_\varepsilon^{(n+1 l-1)}(\zeta)$ 
no more satisfies the second and the third axioms 
(\ref{eqn:cyc}--\ref{eqn:Res}). 

\section{Relations among form factors of the XXZ model}

In this section we shall find further relations among 
$G_\varepsilon^{(nl)}(\zeta)$'s 
and $\tilde{G}_\varepsilon^{(n+1 l-1)}(\zeta)$'s. 

When $N=2n$ we have the following simple relation 
between $G_\varepsilon^{(nn)}(\zeta)$ and 
$G_\varepsilon^{(n-1 n+1)}(\zeta)$: 

\begin{prop}
\begin{equation}
G_\varepsilon^{(n-1 n+1)}(\zeta) = (-1)^n q^{-n-1} 
\pi_\zeta(e_1) G_\varepsilon^{(nn)}(\zeta). 
\label{eqn:nn}
\end{equation}
\label{prop:nn}
\end{prop}

{\sl Proof. }~
Put $n=l$ now. Then it follows 
from (\ref{eqn:Drec}) and (\ref{eqn:e1}) that 
$\Delta_{(n-1 n+1)}^{(1)}$ and 
$(-1)^n q^{n+1}\Delta^{(n-1 n+1)}$ 
have the same recursion relation and the same initial 
condition, and thus the two are the same. 
Since the integral kernel 
$\Psi_\varepsilon^{(mN)}=\Psi_\varepsilon^{(n-1 N)}$ 
is also common 
for $G_\varepsilon^{(n n)}(\zeta)$ and 
$G_\varepsilon^{(n-1 n+1)}(\zeta)$, 
we obtain (\ref{eqn:nn}). ~~~~$\Box$

~

Two homogeneous polynomial 
$\Delta_{(n-1 n+1)}^{(1)}$ and 
$\Delta^{(n-1 n+1)}$ coincide 
up to a constant factor 
as shown in Proposition \ref{prop:nn}. 
The relation 
$\Delta_{(n-1 l+1)}^{(1)}$ and 
$\Delta^{(n-1 l+1)}$ for $n<l$ is not so simple. 
In order to establish the relation, 
let us introduce the symbol $\cong$ as follows: 
We denote 
$A(x_1, \cdots, x_m |z_1, \cdots, z_N)\cong 
 B(x_1, \cdots, x_m |z_1, \cdots, z_N)$ 
when 
$$
\displaystyle 
\prod_{\mu =1}^{m} \oint_{C} \frac{dx_{\mu }}{2\pi i} 
\Psi_{\varepsilon}^{(m N)}(x|\zeta) A(x|z) =
\prod_{\mu =1}^{m} \oint_{C} \frac{dx_{\mu }}{2\pi i} 
\Psi_{\varepsilon}^{(m N)}(x|\zeta) B(x|z). 
$$
Then the following relations hold: 

\begin{prop}
\begin{eqnarray}
&&\Delta_{(n-1 l+1)}^{(1)}(x_1, \cdots, x_n |
z_1, \cdots, z_{n-1}|z_n, \cdots, z_N) 
\label{eqn:updown} \\
&\cong&\displaystyle n(l-n+2)(-1)^{l-n}q^{l+1}(1-q^{-2(l-n)}) 
\prod_{j=1}^{n-1} (x_n -z_jq^{-1}) 
\Delta^{(n-1 l+1)}(x_1, \cdots, x_{n-1} |
z_1, \cdots, z_{n-1}|z_n, \cdots, z_N). 
\nonumber
\end{eqnarray}
\end{prop}

{\sl Proof. }~
The relation (\ref{eqn:updown}) follows from 
the antisymmetry of $x$'s, other than 
the recursion relation and the initial condition 
of $\Delta^{(nl)}$. 

When $l>n=1(N=l+1)$, 
by using (\ref{eqn:e1}) we have 
\begin{equation}
\Delta_{(0 N)}^{(1)}(x_1, \cdots, x_n |
~|z_1, \cdots, z_N) 
=\displaystyle q^{l}\sum_{k=1}^{N} z_k
\frac{\Delta^{(1 l)}(x_1,\cdots,x_n|z_k|
z_{1},\mathop{\hat{\cdots}}^{k},z_N)}
{\prod_{\scriptstyle{i=1}\atop \scriptstyle{i\ne k}}^{N} 
(z_i-z_k)q^{-1}}. 
\label{eqn:e1ini}
\end{equation}
The RHS of (\ref{eqn:e1ini}) is a constant because 
$\mbox{deg} \Delta_{(0 N)}^{(1)} =0$. 
In order to determine this constant, 
we substitute the explicit expression of 
$\Delta^{(1 l)}$ \cite{JKMQ} and put $x_1=0$. 
Then we have 
\begin{equation}
\Delta_{(0 N)}^{(1)}(x_1 |
~|z_1, \cdots, z_N) 
=\displaystyle N(-q)^{l+1}(1-q^{-2(l-1)}). 
\label{eqn:upini}
\end{equation}

When $z_N=z_{n-1}q^{-2}$ in (\ref{eqn:e1}) we have 
\begin{eqnarray}
&&\Delta_{(n-1 l+1)}^{(1)}(x_1, \cdots, x_n |
z_1, \cdots, z_{n-1}|z_n, \cdots, z_N)|_{z_N=z_{n-1}q^{-2}} 
\nonumber \\
&=&(-q)\prod_{\mu=1}^n (x_\mu -z_{n-1}q^{-1}) 
\left\{ \sum_{\mu=1}^{n-1} (-1)^{n+\mu}h(x_\mu)
\Delta_{(n-2 l)}^{(1)}(x_1, \mathop{\hat{\cdots}}^{\mu}, x_{n-1}, 
x_n | z_1, \cdots, z_{n-2}|z_n, \cdots, z_{N-1}) \right. 
\nonumber \\
&+& \left. h(x_n)
\Delta_{(n-2 l)}^{(1)}(x_1, \cdots, x_{n-1} |
z_1, \cdots, z_{n-2}|z_n, \cdots, z_{N-1}) \right\} 
\nonumber \\
&\cong &(-q)\prod_{\mu=1}^n (x_\mu -z_{n-1}q^{-1}) 
\left\{ \sum_{\mu=1}^{n-1} (-1)^{n+\mu}h(x_\mu)
\Delta_{(n-2 l)}^{(1)}(x_1, \mathop{\hat{\cdots}}^{\mu}, x_{n-1}, 
x_n | z_1, \cdots, z_{n-2}|z_n, \cdots, z_{N-1}) \right. 
\nonumber \\
&+&\left. \frac{1}{n-1} \sum_{\mu=1}^{n-1} (-1)^{n+\mu}h(x_\mu)
\Delta_{(n-2 l)}^{(1)}(x_1, \mathop{\hat{\cdots}}^{\mu}, x_{n-1}, 
x_n | z_1, \cdots, z_{n-2}|z_n, \cdots, z_{N-1}) \right\} 
\nonumber \\
&\cong &(1+\frac{1}{n-1}) 
\times (-q) (n-1)(l-n+2)(-1)^{l-n}q^l (1-q^{-2(l-n)}) 
\prod_{\mu=1}^{n} (x_\mu -z_{n-1}q^{-1}) 
\nonumber \\
&\times &\sum_{\mu=1}^{n-1} (-1)^{n+\mu}h(x_\mu)
\prod_{j=1}^{n-2} (x_n -z_{j}q^{-1})
\Delta^{(n-2 l)}(x_1, \mathop{\hat{\cdots}}^{\mu}, x_{n-1} | 
z_1, \cdots, z_{n-2}|z_n, \cdots, z_{N-1}) 
\nonumber \\
&=& n(l-n+2)(-1)^{l-n}q^{l+1} (1-q^{-2(l-n)}) 
\prod_{j=1}^{n-1} (x_n -z_{j}q^{-1})
\prod_{\mu=1}^{n-1} (x_\mu -z_{n-1}q^{-1}) 
\nonumber \\
&\times &\sum_{\mu=1}^{n-1} (-1)^{n-1+\mu}h(x_\mu)
\Delta_{(n-2 l)}^{(1)}(x_1, \mathop{\hat{\cdots}}^{\mu}, x_{n-1} |
z_1, \cdots, z_{n-2}|z_n, \cdots, z_{N-1}) 
\nonumber \\
&=& \left( \mbox{RHS of (\ref{eqn:updown})}\right)|_{z_N=z_{n-1}q^{-2}}, 
\label{eqn:uprec}
\end{eqnarray}
where we use the antisymmetric property with respect of $x$'s, 
and the assumption of the induction, 
in the second and the third equality, respectively. 
Eq. (\ref{eqn:updown}) follows from (\ref{eqn:upini}) 
and (\ref{eqn:uprec}). ~~~~$\Box$

~

Until now, we discuss the case $n\leq l$. 
Let us construct $G_\varepsilon^{(nl)}(\zeta)$ with $n>l$ 
from $G_\varepsilon^{(nn)}(\zeta)$, the spin $0$ sector 
of form factors. 
Define $G_\varepsilon^{(n+1 n-1)}(\zeta)=
\pi_\zeta(f_1)G_\varepsilon^{(nn)}(\zeta)$. Then 
$G_\varepsilon^{(n+1 n-1)}(\zeta)$ also satisfies the three 
axioms with $D=D^{(n-1 n+1)}$. 
By acting $f_1$ successively, we can 
obtain $G_\varepsilon^{(n+k n-k)}(\zeta)$ for $n=1, \cdots, n$, 
just like we construct $G_\varepsilon^{(n-k n+k)}(\zeta)$ from 
$G_\varepsilon^{(nn)}(\zeta)$ by acting $e_1$ successively. 
As for $G_\varepsilon^{(n l)}(\zeta)$ with $n>l$, 
$\pi_\zeta(e_0)G_\varepsilon^{(n l)}(\zeta)=0$ holds. 
The proof is easy if you notice that 
$\pi_\zeta(e_0)G_\varepsilon^{(n n)}(\zeta)=0$ and 
$[e_0, f_1]=0$. 

Note that $G_\varepsilon^{(n l)}(\zeta)$ with $n>l$ is a 
form factor; i.e., $G_\varepsilon^{(n l)}(\zeta)$ satisfies 
the three axioms of form factors with 
$D=D^{(l n)}$. 
You can also show that 
$\tilde{G}_\varepsilon^{(n-1 l+1)}(\zeta):=
\pi_\zeta(e_1)G_\varepsilon^{(n l)}(\zeta)$ 
again satisfies the three axioms with 
$D=D^{(l-1 n+1)}$. 

Let us summarize the relations obtained until now. 

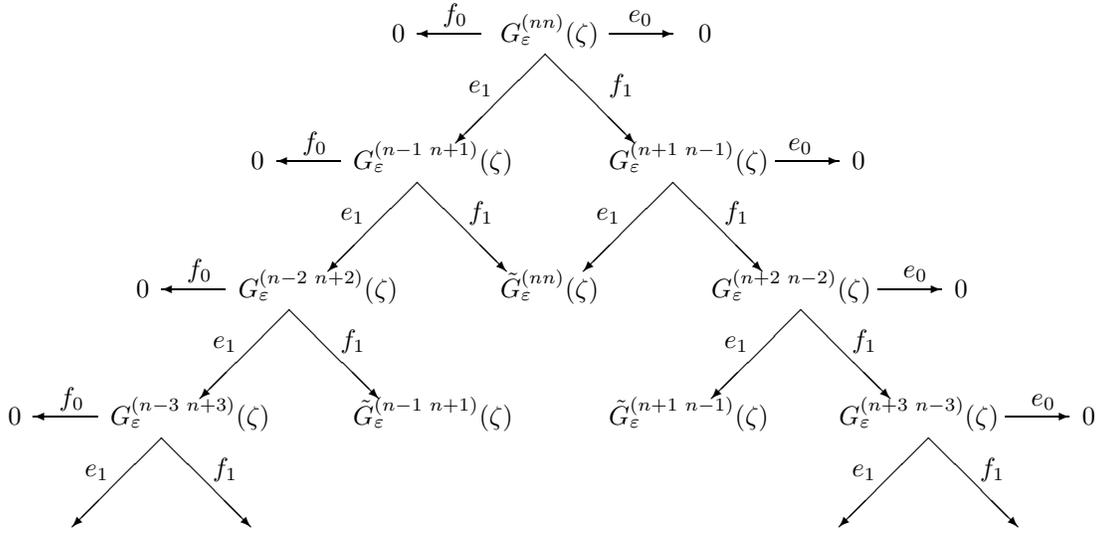
\unitlength 1.7mm
\begin{picture}(90,50)(-5,0)
\put(36.5,40){$G_\varepsilon^{(nn)}(\zeta)$}
\put(35,40.6){\vector(-1,0){5.0}}
\put(32,41.5){$f_0$}
\put(28,40){$0$}
\put(45,40.6){\vector(1,0){5.0}}
\put(46.5,41.5){$e_0$}
\put(52,40){$0$}
\put(40,39){\vector(-1,-1){7.0}}
\put(34,36){$e_1$}
\put(25,30){$G_\varepsilon^{(n-1\,\, n+1)}(\zeta)$}
\put(24,30.6){\vector(-1,0){5.0}}
\put(21,31.5){$f_0$}
\put(17,30){$0$}
\put(40,39){\vector(1,-1){7.0}}
\put(45,36){$f_1$}
\put(45,30){$G_\varepsilon^{(n+1\,\, n-1)}(\zeta)$}
\put(58,30.6){\vector(1,0){5.0}}
\put(59,31.5){$e_0$}
\put(64,30){$0$}
\put(30,29){\vector(-1,-1){7.0}}
\put(24,26){$e_1$}
\put(16,20){$G_\varepsilon^{(n-2\,\, n+2)}(\zeta)$}
\put(15,20.6){\vector(-1,0){5.0}}
\put(12,21.5){$f_0$}
\put(8,20){$0$}
\put(30,29){\vector(1,-1){7.0}}
\put(34,26){$f_1$}
\put(50,29){\vector(-1,-1){7.0}}
\put(44,26){$e_1$}
\put(36.5,20){$\tilde{G}_\varepsilon^{(n n)}(\zeta)$}
\put(50,29){\vector(1,-1){7.0}}
\put(54,26){$f_1$}
\put(53,20){$G_\varepsilon^{(n+2\,\, n-2)}(\zeta)$}
\put(66,20.6){\vector(1,0){5.0}}
\put(68,21.5){$e_0$}
\put(72,20){$0$}
\put(20,19){\vector(-1,-1){7.0}}
\put(14,16){$e_1$}
\put(6,10){$G_\varepsilon^{(n-3\,\, n+3)}(\zeta)$}
\put(5,10.6){\vector(-1,0){5.0}}
\put(2,11.5){$f_0$}
\put(-2,10){$0$}
\put(20,19){\vector(1,-1){7.0}}
\put(24,16){$f_1$}
\put(25,10){$\tilde{G}_\varepsilon^{(n-1\,\, n+1)}(\zeta)$}
\put(60,19){\vector(1,-1){7.0}}
\put(64,16){$f_1$}
\put(63,10){$G_\varepsilon^{(n+3\,\, n-3)}(\zeta)$}
\put(76,10.6){\vector(1,0){5.0}}
\put(78,11.5){$e_0$}
\put(82,10){$0$}
\put(60,19){\vector(-1,-1){7.0}}
\put(54,16){$e_1$}
\put(45,10){$\tilde{G}_\varepsilon^{(n+1\,\, n-1)}(\zeta)$}
\put(10,9){\vector(-1,-1){7.0}}
\put(4,6){$e_1$}
\put(10,9){\vector(1,-1){7.0}}
\put(14,6){$f_1$}
\put(70,9){\vector(-1,-1){7.0}}
\put(64,6){$e_1$}
\put(70,9){\vector(1,-1){7.0}}
\put(74,6){$f_1$}
\end{picture}

~

It is evident from this relationship that 
$G_\varepsilon^{(n-k n+k)}(\zeta)$ and 
$\tilde{G}_\varepsilon^{(n-k n+k)}(\zeta)$ ($-n\leq k \leq n$) 
can be obtained from $G_\varepsilon^{(n n)}(\zeta)$ 
by acting $e_1$ and $f_1$ in an appropriate order. 
We again notice that 
$\pi_\zeta(f_0)G_\varepsilon^{(n n)}(\zeta)=
\pi_\zeta(e_0)G_\varepsilon^{(n n)}(\zeta)=0$. 

We naturally have a form factor 
$F_\varepsilon^{(n n)}(\zeta)$ that belongs to 
$V^{(nn)}$ such that 
$\pi_\zeta(f_1)F_\varepsilon^{(n n)}(\zeta)=
\pi_\zeta(e_1)F_\varepsilon^{(n n)}(\zeta)=0$. 
We can obtain $F_\varepsilon^{(n n)}(\zeta)$ 
from $G_\varepsilon^{(n n)}(\zeta)$ by a 
simple transformation. 

If $G(\zeta)$ solves the three 
axioms of form factors with the diagonal operator $D$, 
then $F(\zeta)=(\sigma^x)^{\otimes N} G(\zeta)$ solves 
them with the diagonal operator $\sigma^x D$. 
Hence $F_\varepsilon^{(ln)}(\zeta)
:=(\sigma^x)^{\otimes N} G_\varepsilon^{(nl)}(\zeta)$ 
and 
$\tilde{F}_\varepsilon^{(ln)}(\zeta)
:=(\sigma^x)^{\otimes N} 
\tilde{G}_\varepsilon^{(nl)}(\zeta)$ 
are also form factors of the XXZ model. 
We can further show that 
$\pi_\zeta(f_1)F_\varepsilon^{(n n)}(\zeta)=
\pi_\zeta(e_1)F_\varepsilon^{(n n)}(\zeta)=0$, and 
$\pi_\zeta(f_1)F_\varepsilon^{(l n)}(\zeta)=0$ for $n<l$, 
$\pi_\zeta(e_1)F_\varepsilon^{(n n)}(\zeta)=0$ for $n>l$. 

Sum up the results obtained in this paper: 
For fixed $n<l$, we find eight form factors 
which belongs to $V^{(nl)}$-sector; i.e., 
$G_\varepsilon^{(nl)}(\zeta)$, 
$\tilde{G}_\varepsilon^{(nl)}(\zeta)$, 
$F_\varepsilon^{(nl)}(\zeta)$ 
and $\tilde{F}_\varepsilon^{(nl)}(\zeta)$, 
where $\varepsilon =\pm$. 
Since we have had 
$G_\varepsilon^{(nl)}(\zeta)$ only when we fix $n<l$ 
at the stage of \cite{KMQ}, we get four times solutions 
of the three axioms of form factor in 
the present work. 

\section{Concluding Remarks} 

In this paper, we have constructed new integral expressions 
of form factors of the XXZ model, by acting 
$U_q \bigl( \widehat{\frak s \frak l _2}\bigr)$ 
to the form factors obtained in \cite{KMQ}. 
The axioms for the form factor 
$G_\varepsilon^{(nl)}(\zeta)$ with 
the diagonal operator $D=D^{(nl)}$ 
reduces those for the form factor 
$\tilde{G}_\varepsilon^{(n+1 l-1)}(\zeta)$ with 
$D=D^{(n-1 l+1)}$ after the action of $f_1$ when 
$n<l$; whereas the axioms for 
$G_\varepsilon^{(nl)}(\zeta)$ with 
$D=D^{(l n)}$ 
reduces those for 
$\tilde{G}_\varepsilon^{(n-1 l+1)}(\zeta)$ with 
$D=D^{(l-1 n+1)}$ after the action of $e_1$ when 
$n>l$. 

The spin $0$ form factor 
$G^{(nn)}(\zeta)$ is a kind of singlet 
because 
$\pi_\zeta(f_0)G^{(n n)}(\zeta)=
\pi_\zeta(e_0)G^{(n n)}(\zeta)=0$. 
In the earlier stage of this work, 
our goal was to decompose the space 
of form factors of the XXZ model into 
infinitely many multiplets of 
$U_q \bigl(\frak s \frak l _2\bigr)$. 
Though $G_\varepsilon^{(nl)}(\zeta)$ 
satisfies three axioms (\ref{eqn:S-symm}--\ref{eqn:Res}) 
and 
$\pi_\zeta(f_0)G_\varepsilon^{(nl)}(\zeta)=0$ when 
$n<l$, 
$\pi_\zeta(e_0)G_\varepsilon^{(nl)}(\zeta)$ no more 
satisfies (\ref{eqn:S-symm}--\ref{eqn:Res}). 
For example, by similar manipulation in (\ref{eqn:actf1}) 
we have 
\begin{equation}
\begin{array}{cl}
&P_{12}\cdots P_{N-1 N}
\pi_{(\zeta_2, \cdots, \zeta_N, \zeta_1 q^{-2})}(e_0) 
G_\varepsilon^{(nl)}(\zeta_2, \cdots, \zeta_N, \zeta_1 q^{-2}) \\
=&\varepsilon r(\zeta_1)D^{(n+1 l-1)}_1 
(q^{2(n-l)}\pi_{\zeta_1}(e_0) \otimes  
\pi_{(\zeta_2, \cdots, \zeta_N)}(1)
+\pi_{\zeta_1}(t_0) \otimes 
\pi_{(\zeta_2, \cdots, \zeta_N)}(e_0))
G_\varepsilon^{(n l)}(\zeta). 
\end{array}
\label{eqn:acte0}
\end{equation}
The RHS of (\ref{eqn:acte0}) reduces to 
$\pi_\zeta(e_0)G_\varepsilon^{(n l)}(\zeta)$ 
up to constant at the limit $q\rightarrow -1$, 
which corresponds the XXX model limit. 

The XXX model has the Yangian 
$Y\bigl(\frak s \frak l _2\bigr)$-symmetry. 
The Yangian $Y\bigl(\frak s \frak l _2\bigr)$ 
is the minimal quantum group which includes  
the universal enveloping algebra 
$U\bigl(\frak s \frak l _2\bigr)$ as a sub-Hopf algebra. 
Since $U\bigl(\frak s \frak l _2\bigr)$ has the 
symmetric coproduct unlike 
$U_q \bigl(\frak s \frak l _2\bigr)$, 
we may be possible  
to decompose the space of form factors 
of the XXX model 
into infinitely many multiplet 
of $U\bigl(\frak s \frak l _2\bigr)$. 
We hope it fruitful to consider the XXX model 
and to find some relations among form factors 
of the model as obtained in this paper. 

\section*{Acknowledgments} The author would like to thank 
M. Jimbo, T. Miwa and A. Nakayashiki for useful discussion. 
He also wishes to thank V. E. Korepin and F. A. Smirnov for 
their interest in this work.

\end{document}